# Selective spatiotemporal mode-locking in multimode Mamyshev oscillators

CHANGFENG WANG[1], JIA LIU[1], YUTAI ZHAO[1], RAN XIA[2,*], WENJUN NI[2], CHUNYONG YANG[2], XIAHUI TANG[1], GANG XU[1] AND PERRY PING SHUM[2,3]

[1]*School of Optical and Electronic Information, Huazhong University of Science and Technology, Wuhan 430074, China*
[2]*Hubei Key Laboratory of Intelligent Wireless Communications; Hubei Engineering Research Center of Intelligent IoT Technology; College of Electronics and Information Engineering (School of Robotics), South-Central Minzu University, Wuhan 430074, China*
[3]*The State Key Laboratory of Optical Fiber and Cable Manufacture Technology, Department of EEE, Southern University of Science and Technology, Shenzhen 518055, China*
**xiaran@mail.scuec.edu.cn*

**Abstract:** Spatiotemporal mode-locking (STML) in multimode fibers introduces spatial degrees of freedom for ultrashort pulse generation, yet flexible manipulation of transverse mode compositions within multimode solitons remains challenging. Here, we propose a spectrum-to-mode mapping strategy in multimode Mamyshev oscillators (MOs) to selectively synchronize specific transverse modes through synergistic cross-phase modulation and offset spectral filtering. Meanwhile, four-wave mixing acts to stabilize intermodal phase relationships, facilitating selective STML characterized by diverse mode compositions with matched group velocities. Spatiotemporal solitons containing the pure $LP_{01}$ mode, degenerate $LP_{11}$ modes, or $LP_{02}$+$LP_{21}$ composite modes are selectively realized and mapped across the parameter space spanned by intracavity gain and filter detuning. Furthermore, intracavity collisions between coexisting mode groups with disparate group velocities are demonstrated to enable dynamic mode selection. Beyond the simulation of six modes in graded-index multimode fibers, selective STML is also demonstrated in MOs based on step-index multimode fibers and fibers supporting more transverse modes, confirming the generality of our approach. This strategy offers a versatile and robust pathway for transverse-mode regulation and provides new insights into full-dimensional structured-light manipulation in multimode ultrafast fiber lasers.

## 1. INTRODUCTION

Multimode fibers (MMFs) support numerous transverse modes and offer substantially larger effective mode areas than single-mode fibers (SMFs) [1–3]. This modal diversity introduces additional spatial degrees of freedom into nonlinear propagation. Spatial overlap among the modes mediates rich intermodal interactions, giving rise to a (3+1)-dimensional framework describing nonlinear spatiotemporal dynamics, such as beam self-cleaning and geometric parametric instability [4–13]. More broadly, the spatial degrees of freedom offered by MMFs enable applications in many fields, including optical imaging [14], mode-division-multiplexed communications [15], and photonic information processing [16–18]. In laser cavities, such intricate intermodal nonlinear interactions can be harnessed to achieve spatiotemporal mode-locking (STML), offering a route to pulse-energy scaling beyond the constraints imposed by the small effective mode areas of SMFs [19]. MMF-based cavities then provide a powerful platform for investigating how additional spatial degrees of freedom reshape complex nonlinear dynamics, including optical turbulence and rogue waves [20,21].

Unlike mode-locking in SMFs, STML requires not only locking the relative phases among longitudinal modes but also synchronizing distinct transverse modes to realize a spatiotemporally coherent state [19]. One primary obstacle for STML originates from modal dispersion, which induces temporal walk-off between different spatial modes. Two mainstream strategies have been developed to address this walk-off challenge. The first strategy integrates multiple intracavity effects, including spectral filtering, saturable absorption, spatial filtering, and spatial coupling, to mitigate modal walk-off and sustain self-consistent periodic evolution of multimode optical fields [22]. In 2017, Wright *et al.* experimentally demonstrated STML for the first time using graded-index multimode fibers (GIMFs) with weak modal dispersion, where additional spatial and spectral filtering was introduced to balance it. They further developed an attractor dissection theoretical model to reveal STML mechanisms and formation dynamics [19,22]. Inspired by this pioneering work, Ding *et al.* validated that saturable absorption alone can compensate for large modal walk-off, removing the strict requirement of weak modal dispersion for STML. Gao *et al.* subsequently proposed a mother-child-mode-coupling theory, suggesting that transverse-mode synchrony relies on spatial coupling to guarantee self-consistent modal evolution [23,24]. Based on these theories, a rich variety of spatiotemporal solitons, soliton molecules, and diverse dynamics have been reported [25–30]. The second mainstream approach employs specialized mode-related hardware to directly compensate differential group walk-off for individual transverse modes. Hu *et al.* utilized mode multiplexers/demultiplexers paired with compensation fibers matched to the group velocity of each transverse mode, achieving synchronization across multiple transverse modes and generating multimode pulses [31–33]. Nevertheless, both approaches come with practical trade-offs: the former demands careful co-optimization of multiple intracavity physical effects, whereas the latter relies on complex mode processing components.

To fully leverage MMFs for pulse energy scaling, Mamyshev oscillators (MOs) have been adopted for STML due to their high tolerance to nonlinear phase shifts and robust energy scalability [34,35]. In 2022, Haig *et al.* reported the first realization of STML within multimode MOs, verifying the feasibility of STML under the Mamyshev regenerative mechanism [36]. Follow-up investigations demonstrated self-starting operation, harmonic mode-locking, and high-power performance in multimode MOs [37–41]. Reported performance includes an average power of 2 W, a pulse energy of 135 nJ, and a pulse duration of 2.2 ps. Furthermore, megawatt-class vortex spatiotemporal solitons have been obtained in multimode MOs by inserting spatial mode-conversion elements such as q-plates, showing great promise for manipulating high-dimensional spatial optical fields of ultrashort pulses [42]. In the studies discussed above, modal dispersion was primarily treated as a detrimental effect to be suppressed or compensated. In contrast, the temporal walk-off and localized pulse overlap can lead to spectral modulation mediated by cross-phase modulation (XPM), effectively turning modal dispersion into a new degree of freedom for mode control.

In this work, we propose an XPM-based spectrum-to-mode mapping mechanism for selective control of the transverse-mode composition of spatiotemporal solitons in multimode MOs. Considering that the mode groups, subsets of modes with well-matched group velocities, undergo partial temporal overlap due to the intergroup group-velocity mismatch, intergroup XPM triggers asymmetric spectral broadening and maps the spatial-mode information into the spectral domain [43–45]. The offset spectral filters then preferentially transmit the mode groups associated with the selected spectral components, thereby completing the spectrum-to-mode mapping. The modes within each selected group subsequently evolve toward a stable state through intracavity gain–loss competition, producing spatiotemporal solitons composed of the fundamental $LP_{01}$ mode, degenerate $LP_{11}$ or $LP_{21}$ modes, or $LP_{02}$+$LP_{21}$ composite modes. Intermodal four-wave mixing (FWM) further establishes phase locking among the constituent modes, yielding ultrafast Laguerre–Gaussian (LG) vortex solitons. Using this MO architecture, we demonstrate selective STML with distinct modal compositions in both six-mode GIMFs and step-index multimode fibers (SIMFs) and further validate the proposed framework in a fiber supporting 15 modes. These results demonstrate that modal dispersion can serve as a facilitating resource for mode-resolved spatiotemporal-field engineering in multimode ultrafast lasers.

## 2. THEORETICAL MODEL

Figure 1 illustrates the physical mechanism underlying selective STML. We contrast the selective STML enabled by XPM and offset filtering in the MO against conventional STML, highlighting that while the latter still experiences considerable temporal walk-off [22,24], our approach effectively avoids significant walk-off once stable mode-locking is established. Specifically, mode selection is realized by exploiting intergroup XPM to induce spectral shifts between modes exhibiting group-velocity mismatch [see Fig. 1(e)]. The offset filters then convert these spectral differences into mode-dependent transmission, realizing the spectrum-to-mode mapping. The selective STML state is finally achieved via self-consistent dissipative spatiotemporal dynamics within the MO cavity.

We first quantify the proposed spectrum-to-mode mapping by simulating multimode pulse propagation followed by offset spectral filtering. Propagation in the MMF is described by the simplified generalized multimode nonlinear Schrödinger equation (GMMNLSE) given in Eq. (A2) of Appendix A. This reduced model includes propagation-constant mismatch, modal dispersion, and corresponding nonlinear terms for self-phase modulation (SPM) and XPM. The instantaneous-frequency shift $\delta\omega_p$ imposed on mode $p$ by SPM and XPM is given by:

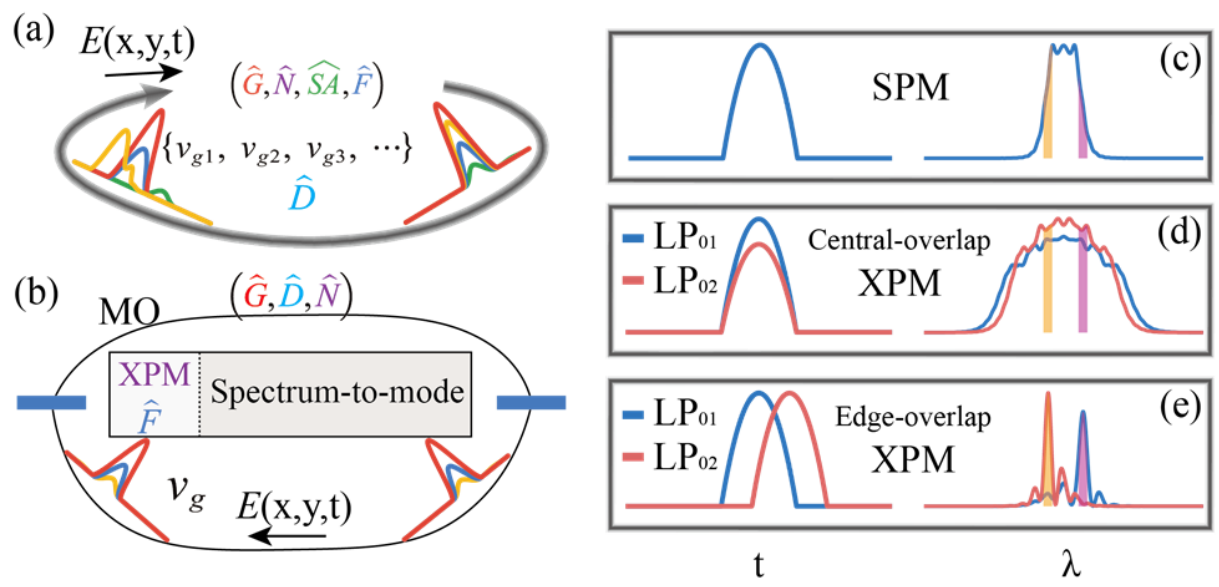

Fig. 1. Schematic and underlying physical mechanism of selective STML. (a) Schematic of conventional STML, where operators $\hat{G}$, $\hat{N}$, $\widehat{SA}$, $\hat{F}$, and $\hat{D}$ denote gain, nonlinearity, saturable absorption, filtering (both spatial and spectral), and linear (dispersion and loss) effects, respectively. This process relies on the synergistic interplay of these effects to balance the walk-off among modes with different group velocities. (b) Schematic of the proposed selective STML, which couples XPM-induced spectral shifting with offset spectral filters in the MO to establish a mode-locked state supported by modes with closely matched group velocities. (c)–(e) Temporal and spectral profiles for (c) SPM, (d) intergroup XPM with coincident pulse centers (central-overlap XPM), and (e) intergroup XPM with partial temporal overlap at the pulse edges (edge-overlap XPM). The spectra in (c) and (d) undergo approximately symmetric broadening, whereas those in (e) exhibit asymmetric broadening and mode-dependent spectral shifts. The dashed vertical bands indicate the passbands of spectral filters.

$$\delta\omega_p(L,\tau) = -\frac{n_2\omega_0}{c}\int_0^L \left[ S^K_{pppp}\frac{\partial |A_p(z,\tau)|^2}{\partial\tau} + \sum_{n\neq p}^{N}(S^K_{ppnn}+S^K_{pnpn})\frac{\partial |A_n(z,\tau+\delta\beta^{(1)}_{p,n}z)|^2}{\partial\tau}\right]dz \tag{1}$$

where $n_2$ is the nonlinear refractive index for silica glass fibers, $\omega_0$ is the center angular frequency, and $c$ is the speed of light in vacuum. $A_m(z,\tau)$ denotes the field envelope of mode $m$. $S^K_{pppp}$ represents the nonlinear coupling coefficient of SPM, while $S^K_{pnpn}$ and $S^K_{ppnn}$ describe XPM. The term $\delta\beta^{(1)}_{p,n} = \beta^{(1)}_p - \beta^{(1)}_n$ describes the modal dispersion, i.e., the group-velocity mismatch between modes $p$ and $n$.

Here, we consider a 50-cm-long GIMF with a 30-μm core diameter and an NA of 0.08, which supports six transverse modes at the carrier wavelength of 1030 nm. Based on different group velocities, these modes can be categorized into two distinct groups: a slow mode group ($LP_{01}$, $LP_{11a}$, and $LP_{11b}$) and a fast mode group ($LP_{02}$, $LP_{21a}$, and $LP_{21b}$) [Fig. 10], forming a slow and a fast pulse in the time domain, respectively. Guided by the gradient-dependent frequency shift in Eq. (1), we scan the initial temporal separation of the pulses to control the extent and location of pulse overlap, thereby regulating the intergroup XPM. Given that the modes within each group have nearly identical group velocities and undergo similar spectral evolution, $LP_{11b}$ and $LP_{21b}$ are selected as representative modes of the slow and fast pulses, respectively.

After propagation, spectral filters with fixed passbands are applied to evaluate how the spectral differences translate into preferential modal transmission. Figure 2(a) plots the resulting $LP_{11b}$-to-$LP_{21b}$ peak-power ratio as a function of the initial temporal separation. When the fast and slow pulses overlap at their temporal edges [Figs. 2(c) and 2(e)], intergroup XPM produces asymmetric spectral broadening, in contrast to the approximately symmetric broadening observed in Figs. 2(b) and 2(d) [45,46]. During edge overlap, the pulse overlapping on its leading edge is blue-shifted, whereas the pulse overlapping on its trailing edge is red-shifted, as predicted by Eq. (1). A spectrally offset filter then imposes different transmission losses on the two mode groups, preferentially transmitting the mode group whose shifted spectrum overlaps more strongly with its passband.

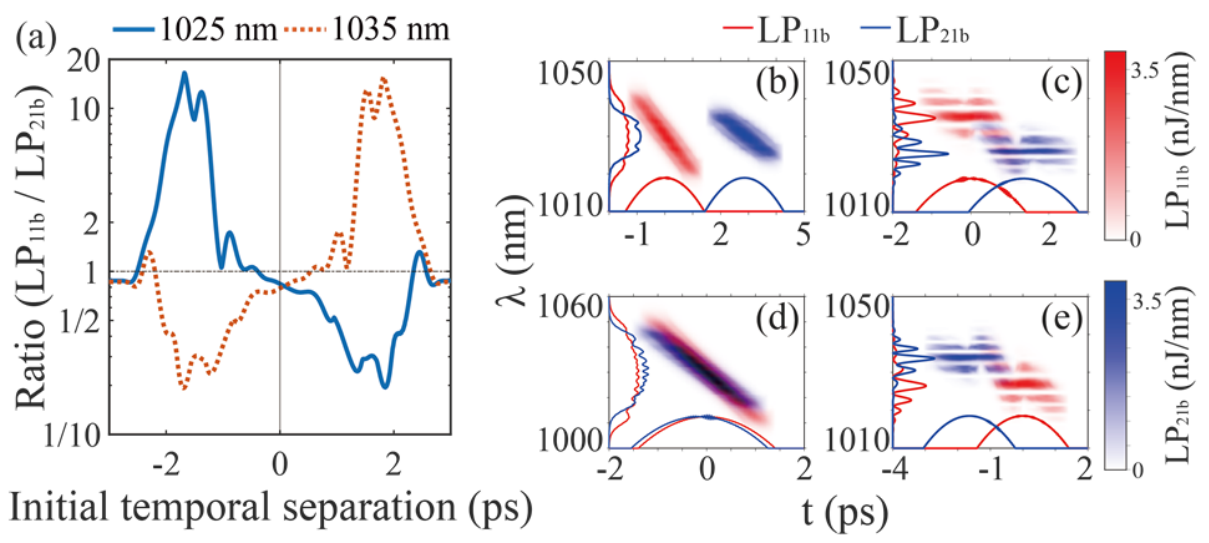

Fig. 2. Spectrum-to-mode mapping. (a) $LP_{11b}$-to-$LP_{21b}$ peak-power ratio after propagation and spectral filtering, plotted as a function of the initial temporal separation. The blue solid and orange dotted curves correspond to Gaussian filters centered at 1025 and 1035 nm, respectively, each with a 3-dB bandwidth of 2 nm. (b)–(e) Spectrograms of $LP_{11b}$ and $LP_{21b}$ at the fiber output, for initial temporal separations of 3, 1.5, 0, and −1.5 ps, respectively. The insets show the corresponding temporal and spectral profiles.

## 3. RESULTS AND DISCUSSION

The schematic of the multimode MO designed to enable selective STML is shown in Fig. 3. The numerical simulations are performed using the GMMNLSE given in Eq. (A1) [see Appendix A], with the linear parameters and nonlinear coupling coefficients obtained by solving the waveguide eigenmode equations [see Fig. 10]. The small-signal gain of the ytterbium-doped graded-index fiber (YDF) is modeled using a Gaussian spectral profile with a FWHM of 40 nm and a peak coefficient $g_0$ of 20 dB/m, whereas no gain is applied in the passive graded-index fiber (GDF) sections. The output-coupling ratio of the intracavity output coupler (OC) is set to 0.5. The intracavity spectral filters, each featuring a Gaussian transmission profile with a 3-dB bandwidth of 2 nm, have central wavelengths $\lambda_1$ (Filter1) and $\lambda_2$ (Filter2) symmetrically detuned around 1030 nm, where the wavelength detuning is defined as $\delta\lambda = \lambda_1 - \lambda_2$ . To implement the spectrum-to-mode mapping described above in the MO cavity, a relatively large filter separation $|\delta\lambda|$ is adopted, allowing the offset passbands to discriminate between the XPM-shifted spectra of different mode groups. This filtering condition also strongly suppresses continuous-wave (CW) oscillation and hinders self-starting [47], so an external seed pulse is injected to initiate oscillation.

To investigate how spectrum-to-mode mapping leads to selective STML, we first consider a spectrally pre-conditioned seed to illustrate the direct mode selection. Before cavity injection, the seed propagates through an MMF, where modal walk-off and intergroup XPM generate red- and blue-shifting of the fast and slow mode groups, respectively [see Appendix C]. Filter1 then converts these spectral differences into differential loss: the short-wavelength (long-wavelength) passband preferentially transmits the slow (fast) mode group. The surviving field then undergoes gain–loss competition and nonlinear evolution, which jointly determine the final transverse-mode composition and establish a stable selective STML state in the MO.

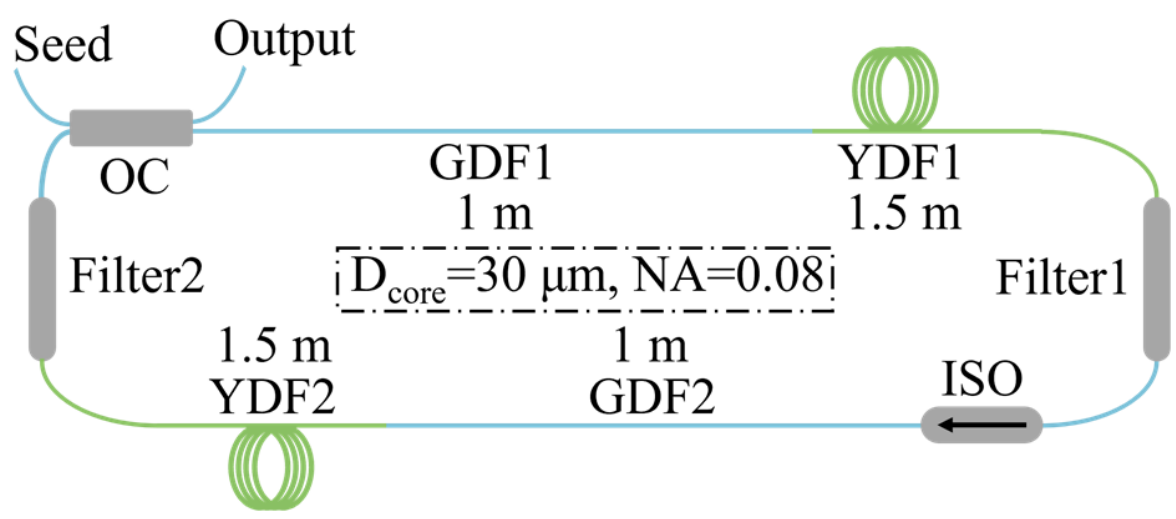

Fig. 3. Schematic of the multimode MO. The cavity has two arms, each containing 1.5 m of YDF, 1 m of passive GDF, and a spectral filter. All fiber sections are MMFs with a core diameter of 30 μm and an NA of 0.08. ISO: isolator.

To identify the specific modes that can be locked in the MO, we systematically map the resulting mode-selection regimes across the parameter space spanned by $\delta\lambda$ and the saturation fluence $F_{\mathrm{sat}}$. We first scan $F_{\mathrm{sat}}$ and $\delta\lambda$ in steps of 5 J/m$^2$ and 1 nm, respectively, then refine the steps to 1 J/m$^2$ and 0.2 nm near the boundaries between distinct mode-locking states. Each parameter point is classified according to the final stable intracavity optical field after sufficient roundtrip evolution. The regions distinguished by different colors in Fig. 4(a) indicate the mode group selected by Filter1, whereas the discrete symbols identify the final mode composition of the pulse. The representative spectral and temporal evolution in Figs. 4(b-e) illustrates the initial spectrum-to-mode mapping imposed on the seed by Filter1.

In the "None" region (gray), spectral broadening is insufficient for the pulse spectrum to overlap with the filter passband at large $|\delta\lambda|$. Stable oscillation therefore fails to develop, as denoted by the black squares. In the "Slow" region (green), corresponding to $\delta\lambda < 0$, the fast mode group decays rapidly within the first few cavity roundtrips, while the slow group components survive and continue to compete within the cavity, as illustrated in Figs. 4(b) and 4(d). Subsequent intragroup competition leads to either a pure $LP_{01}$ mode soliton (red star) or an $LP_{11}$ degenerate-mode soliton (blue circle). At a fixed $F_{\mathrm{sat}}$, the $LP_{11}$ soliton occurs predominantly at smaller $|\delta\lambda|$, whereas the $LP_{01}$ soliton prevails at larger $|\delta\lambda|$. Because the SPM overlap coefficient $S^{K}_{pppp}$ is larger for $LP_{01}$ than for $LP_{11}$ ($5.10 \times 10^9$ m$^{-2}$ versus $3.82 \times 10^9$ m$^{-2}$, Fig. 10), the $LP_{01}$ mode accumulates a larger nonlinear phase shift and undergoes stronger spectral broadening at comparable modal energies, giving the $LP_{01}$ mode an advantage in the modal competition at larger detunings. At smaller detuning magnitudes, however, the broader $LP_{01}$ spectrum has a lower spectral intensity within the filter passband, allowing the $LP_{11}$ degenerate modes to prevail. These effects divide the "Slow" region into two distinct mode-selection subregions. The inset in Fig. 4(d) shows the evolution of the $LP_{01}$ energy fraction at $F_{\mathrm{sat}} = 50$ J/m$^2$ for two filter detunings, $\delta\lambda = -7$ nm and $-9$ nm, illustrating convergence toward the $LP_{11}$ and $LP_{01}$ solitons, respectively.

In the "Fast" region (blue, $\delta\lambda > 0$), Filter1 selectively transmits the fast mode group, as illustrated in Figs. 4(c)

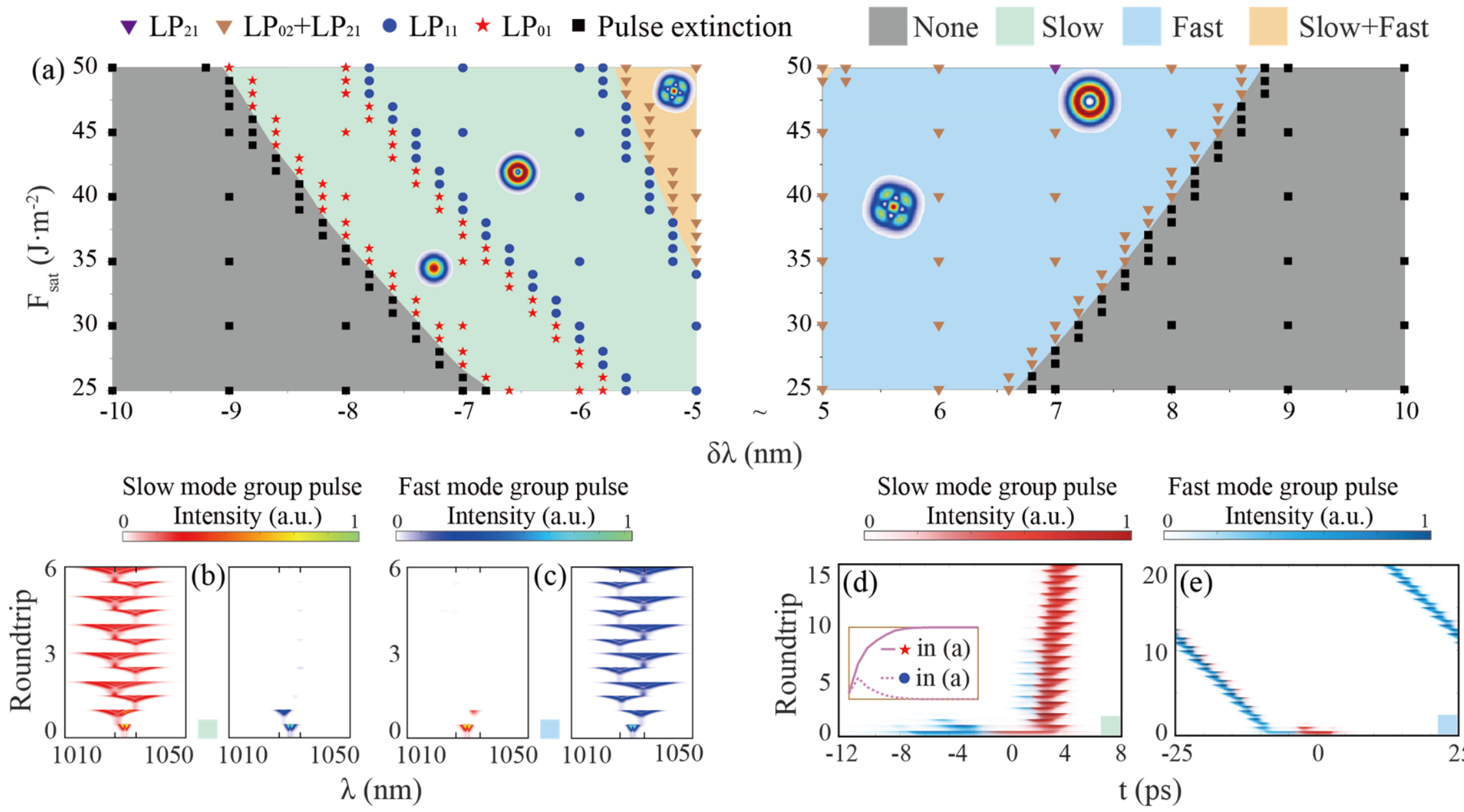

Fig. 4. Mode selection and intracavity evolution governed by spectrum-to-mode mapping in the MO. (a) Parameter space of modal dynamics. Based on the action of Filter1 on the seed pulse, the parameter space is divided into four regimes: no-pulse (None), slow-pulse-selected (Slow), fast-pulse-selected (Fast), and fast–slow pulse coexistence (Slow+Fast), indicated by gray, green, blue, and brown backgrounds, respectively. For $-5\,nm < \delta\lambda < 5\,nm$, small detuning magnitudes promote pulse splitting for the same modal composition. Because such states fall outside the scope of the present study, this parameter range is not systematically analyzed. Symbols denote the final states reached after sufficient intracavity evolution: stable spatiotemporal solitons formed by the degenerate $LP_{21}$ modes (purple triangles), the $LP_{02}$+$LP_{21}$ modes (brown triangles), the degenerate $LP_{11}$ modes (blue circles), or the $LP_{01}$ mode (red stars), and pulse extinction (black squares). (b,c) Spectral and (d,e) temporal evolution of the intracavity field over roundtrips. (b,d) and (c,e) correspond to the green and blue regions in (a), respectively. The inset in (d) shows the energy fraction of the $LP_{01}$ mode versus roundtrips for two cases in the green region that converge to the $LP_{11}$ degenerate-mode soliton and the $LP_{01}$ mode soliton, respectively.

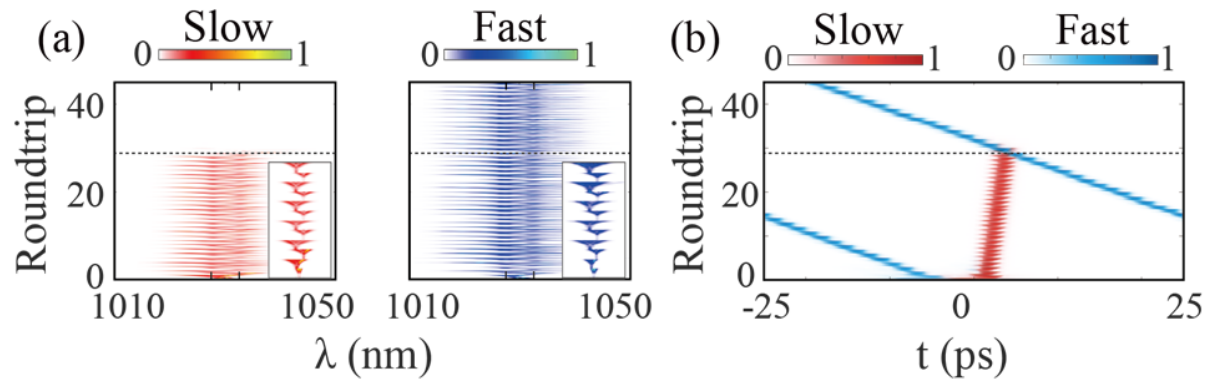

Fig. 5. (a) Spectral and (b) temporal evolution of the intracavity field over roundtrips at $g_0 = 20$ dB/m, $F_{\mathrm{sat}} = 50$ J/m$^2$, and $\delta\lambda = -5$ nm. The insets in (a) show the spectral evolution of the first six roundtrips, during which the slow and fast mode groups coexist and develop stable spectral oscillations.

and 4(e). This region predominantly supports an $LP_{02}$+$LP_{21}$ composite-mode soliton, and an $LP_{21}$ degenerate-mode soliton can also form at specific parameter points (e.g., $F_{\mathrm{sat}} = 50$ J/m$^2$ and $\delta\lambda = 7$ nm). Together, the "Slow" and "Fast" regions delineate mode-selective operational domains, offering a straightforward route to switch transverse modes and deliver the corresponding ultrashort pulses via simple tuning of the MO.

When $|\delta\lambda|$ is relatively small, Filter1 fails to selectively suppress either mode group, allowing the fast and slow pulses to coexist during the early cavity evolution, denoted as the "Slow+Fast" region (brown). Both pulse components exhibit comparable cyclical dynamics, characterized by alternating stages of nonlinear spectral broadening and subsequent bandpass filtering per roundtrip [Fig. 5(a)]. Meanwhile, modal dispersion continuously changes their relative temporal separation [Fig. 5(b)], eventually driving the two pulses into collision. Upon temporal overlap, intergroup XPM again induces asymmetric spectral broadening. Subsequent filtering converts the resulting spectral asymmetry into differential loss, thereby breaking the coexistence of the two pulses and preferentially retaining one of them. Spectrum-to-mode mapping can therefore be initiated by not only the spectrally pre-conditioned seed but also intracavity pulse collisions.

To systematically investigate collision dynamics over an expanded coexistence domain, $g_0$ is elevated from 20 to 30 dB/m, with all other cavity parameters and the protocol for seed preconditioning kept unchanged. The two-parameter analysis over $F_{\mathrm{sat}}$ and $\delta\lambda$ is then repeated, yielding the parameter map shown in Fig. 6(a), which retains the "Slow" and "Fast" regions observed in Fig. 4(a), but exhibits a substantially larger "Slow+Fast" region. Since modal dispersion enables the fast pulse to catch up with the slow pulse, they undergo an edge-overlap collision analogous to the configuration depicted in Fig. 2(c). Intergroup XPM redshifts the slow pulse and blueshifts the fast pulse, allowing the subsequent long- or short-wavelength filter to selectively transmit the slow or fast mode group, respectively. Consequently, the position of collision determines which filter acts on the collision-modified spectra and thereby controls the selection of a mode group.

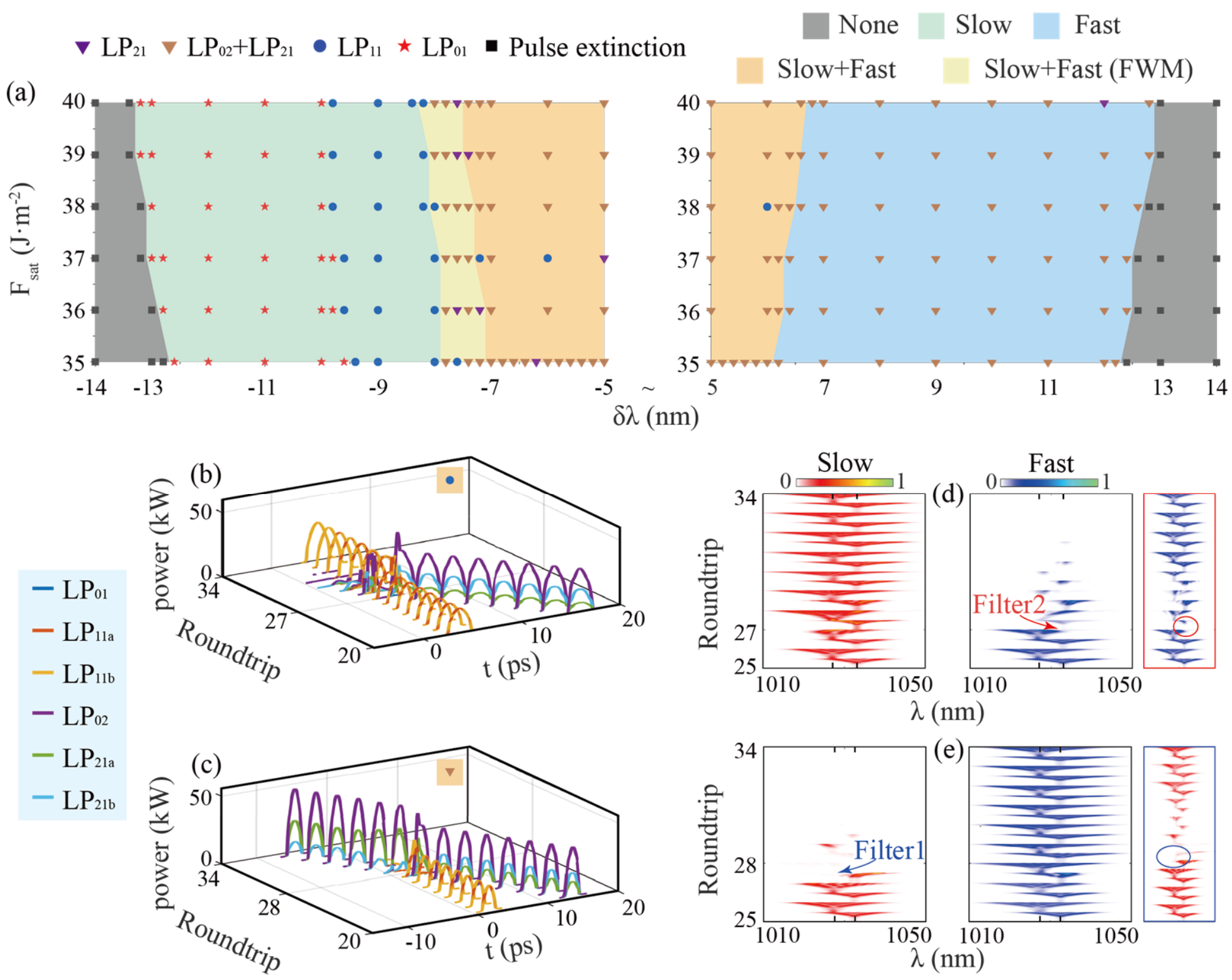

Fig. 6. Mode selection and intracavity collision dynamics in the MO at $g_0 = 30$ dB/m. (a) Parameter space of modal dynamics. Symbols and background colors follow the conventions in Fig. 4(a), with an additional yellow background region indicating the coexistence of fast and slow pulses mediated by FWM. (b,c) Three-dimensional plots of the temporal evolution of the intracavity field over roundtrips. (d,e) Spectral evolution of the intracavity field over roundtrips. (b,d) and (c,e) correspond to pulse collisions before Filter2 and Filter1, respectively, leading to the selection of the slow and fast mode groups for $\delta\lambda < 0$. The insets to the right of panels (d) and (e) illustrate cases in which spectral filtering does not select a single mode group after a collision.

Figures 6(b)–6(e) illustrate these two selection pathways. For the negative filter detuning considered here, Filter1 is centered on the short-wavelength side, while Filter2 is centered on the long-wavelength side. At $F_{\mathrm{sat}} = 37$ J/m$^2$ and $\delta\lambda = -6$ nm, the principal collision occurs during the 27th roundtrip just before Filter2, after which the cavity converges to an $LP_{11}$ degenerate-mode soliton [Figs. 6(b) and 6(d)]. By contrast, at $F_{\mathrm{sat}} = 40$ J/m$^2$ and $\delta\lambda = -5$ nm, the collision occurs during the 28th roundtrip before Filter1, resulting in an $LP_{02}$+$LP_{21}$ composite-mode soliton [Figs. 6(c) and 6(e)]. Thus, the long-wavelength filter encountered immediately after the collision selects slow modes, while the short-wavelength filter selects fast modes. These contrasting outcomes are consistent with the spectrum-to-mode mapping established above.

A single collision followed by spectral filtering, however, does not always complete mode-group selection. When the XPM-induced spectral shift is insufficient for the subsequent filter to introduce adequate differential loss, the fast and slow pulses remain in transient coexistence [insets on the right of Figs. 6(d) and 6(e)]. Additional collisions are therefore required until only modes with similar group velocities remain in the cavity, culminating in a stable mode-locked state.

While the mechanisms discussed above focus on the selective suppression and extinction of a specific mode group, the regeneration of a mode group can also occur. Specifically, within the "Slow+Fast (FWM)" region (yellow) of the parameter space, the fast pulse is regenerated via intergroup FWM following its initial extinction. For instance, at $F_{\mathrm{sat}} = 37$ J/m$^2$ and $\delta\lambda = -8$ nm [Fig. 7(a)], Filter1 strongly suppresses the fast-mode components of the injected seed, but subsequent FWM-mediated energy transfer restores the fast pulse and enables coexistence with the slow pulse. Efficient FWM requires the following energy-conservation and phase-matching conditions [1]:

$$\begin{aligned} \omega_1 + \omega_2 - \omega_3 - \omega_4 &= 0 \\ \beta_3(\omega_3) + \beta_4(\omega_4) - \beta_1(\omega_1) - \beta_2(\omega_2) + \Delta\beta_{NL} &= 0 \end{aligned} \quad (2)$$

where $\Delta\beta_{NL}$ denotes the nonlinear contribution to the propagation-constant mismatch. Satisfying these conditions permits efficient intermodal energy exchange. In the present case, the simulations show a net transfer from the slow-mode components to the fast-mode components, parametrically regenerating the fast pulse and allowing the two pulses to coexist. These states subsequently undergo collision dynamics analogous to those in Fig. 6, leading to selective STML.

Beyond transferring energy between the fast and slow mode groups, FWM provides phase-sensitive coupling that can establish specific phase relationships among the modes within the same mode group. Figs. 7(b-d) and 7(e-g) show the evolution of the modal energy fractions and intermodal phase differences, respectively, for three representative dynamical scenarios. In the first scenario [Figs. 7(b) and 7(e)], corresponding to $F_{\mathrm{sat}} = 40$ J/m$^2$ and $\delta\lambda = -8$ nm, the cavity yields an $LP_{02}$+$LP_{21}$ composite-mode soliton. The relative phase between $LP_{02}$ and $LP_{21a}$ modes locks at $|\Delta\phi| = \pi/2$, while $\Delta\phi$ between the two $LP_{21}$ degenerate modes is zero. These stationary phase relationships demonstrate coherent coupling among the constituent modes of the composite pulse.

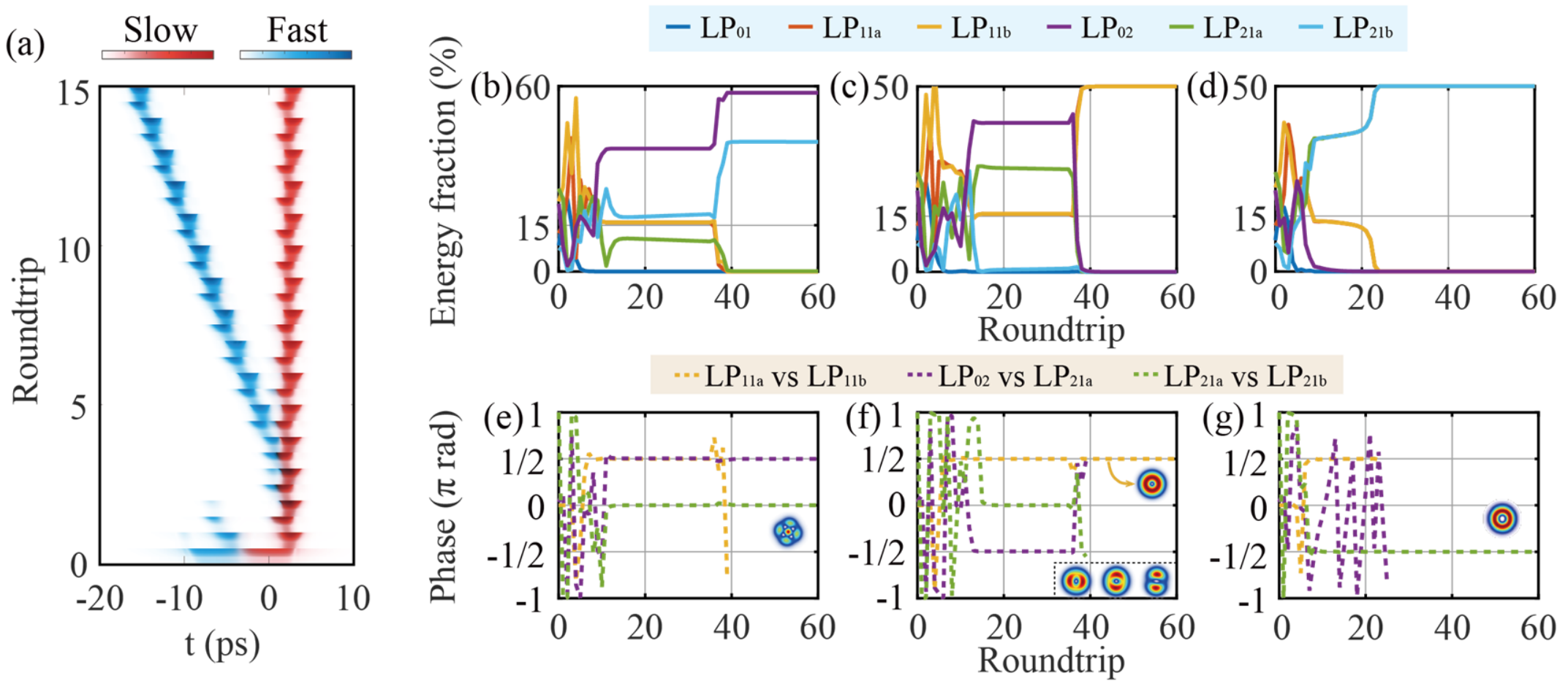

Fig. 7. (a) Temporal evolution of the intracavity field over roundtrips in the Slow+Fast (FWM) region. The fast pulse is regenerated from the slow pulse. (b-d) Evolution of the modal energy fractions, with colors identifying the modes as indicated in the legend. (e-g) Evolution of the intermodal phase differences for the mode pairs specified in the legend. (b,e), (c,f), and (d,g) correspond to the formation of $LP_{02}$+$LP_{21}$ composite-mode, $LP_{11}$ degenerate-mode, and $LP_{21}$ degenerate-mode solitons, respectively. The insets show the corresponding transverse intensity profiles, while those enclosed by dashed lines in (f) show $LP_{11}$ degenerate-mode solitons obtained when FWM is neglected.

In the second scenario, at $F_{\text{sat}} = 40$ J/m$^2$ and $\delta\lambda = -8.2$ nm, the surviving $LP_{11}$ degenerate-mode soliton exhibits equal-amplitude components with a relative phase of $|\Delta\phi| = \pi/2$ [Figs. 7(c) and 7(f)]. The coherent superposition of these components yields an ultrafast $LG_{01}$ vortex pulse with an azimuthally uniform annular intensity profile. To isolate the role of FWM, the same simulation is repeated with the FWM terms removed. Although the two $LP_{11}$ components retain equal amplitudes and their relative phase still approaches a constant, $|\Delta\phi|$ is no longer fixed at $\pi/2$. The resulting transverse intensity profile is therefore annular but azimuthally nonuniform [insets of Fig. 7(f)]. This comparison shows that FWM selects a specific relative phase, rather than merely maintaining a constant one. Together with the composite-mode results, this finding helps explain why states with the same modal composition exhibit similar phase relationships and transverse-field structures under different operating conditions [see Fig. 4].

The third scenario illustrates a distinct selection pathway in which an $LP_{21}$ degenerate-mode soliton forms without a pronounced intergroup collision. At $F_{\text{sat}} = 40$ J/m$^2$ and $\delta\lambda = -7.6$ nm [Figs. 7(d) and 7(g)], the $LP_{02}$ component does not establish a stable phase relationship with the $LP_{21}$ components and is progressively suppressed, leaving the fast pulse composed solely of the $LP_{21a}$ and $LP_{21b}$ modes. During the transient coexistence stage, the slow $LP_{11}$ pulse carries substantially less energy than in the first scenario and is suppressed through gain–loss competition before a pronounced intergroup collision occurs. The surviving $LP_{21a}$ and $LP_{21b}$ components subsequently evolve toward a stable state with equal amplitudes [Fig. 7(d)] and a locked relative phase of $|\Delta\phi| = \pi/2$ [Fig. 7(g)], forming an ultrafast $LG_{02}$ vortex pulse.

The preceding analysis demonstrates that FWM enables intermodal energy transfer and locks the relative phases of the surviving transverse modes, establishing spatiotemporal coherence. Meanwhile, it should be noted that within the "Slow+Fast (FWM)" region in Fig. 6, solitons composed of different transverse mode groups are obtained. This diversity arises because the ultimate mode selection in this region is governed by the intracavity collision process, whose location cannot be deterministically controlled via $F_{\text{sat}}$ and $\delta\lambda$. Hence, achieving more precise control over the collision dynamics warrants further exploration. As illustrated in Figs. 7(b), 7(c), 7(e), and 7(f), the energy fractions of the fast and slow pulses reach a steady state prior to collision, accompanied by stabilized intermodal phase relationships, yielding a constant roundtrip temporal walk-off. Consequently, tailoring the initial temporal separation between the coexisting fast and slow pulses provides a deterministic means to govern their intracavity collision coordinate.

In Fig. 8(a), we re-inject the optical field of the stably coexisting fast and slow pulses as an external seed into the MO ($g_0 = 30$ dB/m, $F_{\text{sat}} = 50$ J/m$^2$, and $\delta\lambda = -8$ nm) and systematically scan the initial temporal separation from 3 to 6.2 ps in steps of 0.01 ps, revealing two distinct mode-selection windows demarcated in blue and red. The blue window corresponds to collisions occurring immediately before Filter1, which yield an $LP_{02}$+$LP_{21}$ composite-mode soliton with a total energy of 189 nJ. Conversely, the red window corresponds to collisions occurring before Filter2, producing an $LP_{11}$ degenerate-mode soliton with a total energy of 131 nJ. These selection windows alternate with a period of approximately 1.5 ps, matching the 1.51 ps intergroup walk-off accumulated per roundtrip. The distinct energy levels are related to the mode-dependent effective areas. Because the effective modal area $A_{eff}$ scales inversely with the SPM overlap coefficient $S^{K}_{pppp}$ [values provided in Fig. 10(b)], modes with larger effective areas sustain higher saturation energies at a fixed $F_{\text{sat}}$, stabilizing at higher pulse energies.

Between these selection windows, inadequate XPM-induced spectral shifts yield insufficient differential loss across the filter, resulting in the coexistence of the fast and slow pulses with a total energy of 216 nJ. It should be noted that similar coexistence can also occur within the selection windows. This behavior arises because the multi-peak spectral structure [Fig. 2] can cause residual spectral components of the disfavored mode group to leak across the passband of the filter. Additionally, when the fast pulse survives the collision, it can be composed solely of the degenerate $LP_{21}$ modes, yielding a slightly lower pulse energy than its counterpart comprising both $LP_{02}$ and $LP_{21}$ modes.

Having clarified the relationship between initial pulse separation and collision-mediated mode selection in the GIMF cavity, we next demonstrate a practical route to manipulate this separation by varying the length of the preprocessing fiber through which the seed pulse propagates before entering the cavity (Fig. 11). To assess this control scheme under altered modal-dispersion landscapes, we retain the MO configuration (Fig. 3) and seed-generation scheme (Fig. 11), but replace the fibers with six-mode SIMFs (20 μm core diameter, NA = 0.08).

The preprocessing length $L_{\text{pre}}$ provides a practical means to tune the relative pulse timing via intermodal walk-off prior to cavity injection, with the differential delay governed by $\Delta t_{pre} = \delta\beta^{(1)}_{p,n} L_{\text{pre}}$. $LP_{11}$ and $LP_{21}$ degenerate modes constitute the slow and fast pulses, respectively, consistent with the modal dispersion of the SIMF [Fig. 10]. As $L_{\text{pre}}$ increases, the pre-collision temporal separation decreases linearly with a slope of $-0.89$ ps/m [orange squares in Fig. 8(b)], in excellent agreement with the group-velocity mismatch between the $LP_{21}$ and $LP_{11}$ modes. This linear relationship establishes a direct, deterministic mapping between the external preprocessing fiber length $L_{\mathrm{pre}}$ and the intracavity collision coordinate.

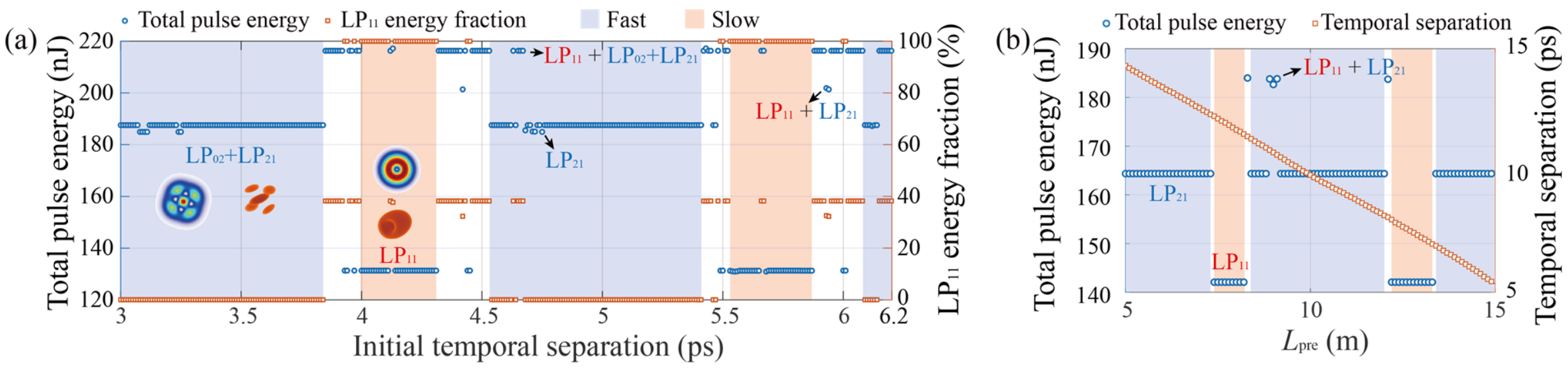

Fig. 8. Control of pulse-collision outcomes in the MO by adjusting the temporal separation between fast and slow pulses to set the collision position. (a) Total pulse energy ($E_{tot} = \sum_{p=1}^{6} \int |A_p(z,t)|^2 dt$, blue) and energy fraction of the LP$_{11}$ degenerate-mode group (orange) after a single collision and subsequent evolution to a stable state, plotted against the initial temporal separation in an MO comprising GIMF (30-µm-core, NA = 0.08). Distinct energy plateaus correspond to different modal compositions, with representative transverse field profiles and three-dimensional soliton structures shown in the insets. (b) Pre-collision temporal separation between the pulses (orange) and total pulse energy after stabilization (blue) versus the length $L_{\mathrm{pre}}$ of SIMF (20-µm-core, NA = 0.08) traversed by the seed pulse before entering an MO comprising SIMF. The temporal separation varies linearly with $L_{\mathrm{pre}}$, allowing the collision position to be controlled.

At $g_0 = 30$ dB/m, $F_{\mathrm{sat}} = 50$ J/m$^2$, and $\delta\lambda = -9$ nm, varying $L_{\mathrm{pre}}$ shifts the intracavity collision position and thereby controls the final modal composition, as indicated by the blue circles in Fig. 8(b). The total pulse energy exhibits three distinct plateaus at 142, 164, and 184 nJ, corresponding to the LP$_{11}$ degenerate-mode soliton, the LP$_{21}$ degenerate-mode soliton, and the coexistence state of the fast and slow pulses, respectively. These states recur as the relative pulse separation is varied, with a period of approximately 4.45 ps, matching the walk-off accumulated over one roundtrip in the MO cavity. Specifically, an LP$_{11}$ (LP$_{21}$) degenerate-mode soliton is selected when the collision occurs prior to Filter2 (Filter1).

We further examine whether the spectrum-to-mode mapping framework remains scalable to systems supporting a larger number of modes. To this end, we numerically investigate an MO based on 15-mode GIMFs. Given the substantial computational overhead, the exploration focuses on representative mode-selection regimes under XPM-based seed preconditioning. Figure 9 shows two stable composite-mode solitons obtained at $g_0 = 30$ dB/m and $F_{\mathrm{sat}} = 50$ J/m$^2$. At $\delta\lambda = -9$ nm [Fig. 9(a)], the cavity selects the LP$_{02}$+LP$_{21}$ composite-mode group, whereas at $\delta\lambda = 9$ nm [Fig. 9(b)], it isolates the LP$_{12}$+LP$_{31}$ composite-mode group. Within each group, the constituent modes possess closely matched group velocities, while intergroup walk-off remains pronounced. Spectrum-to-mode mapping thus effectively isolates the targeted group to achieve selective STML, confirming the versatility of this framework in multimode systems with larger mode counts.

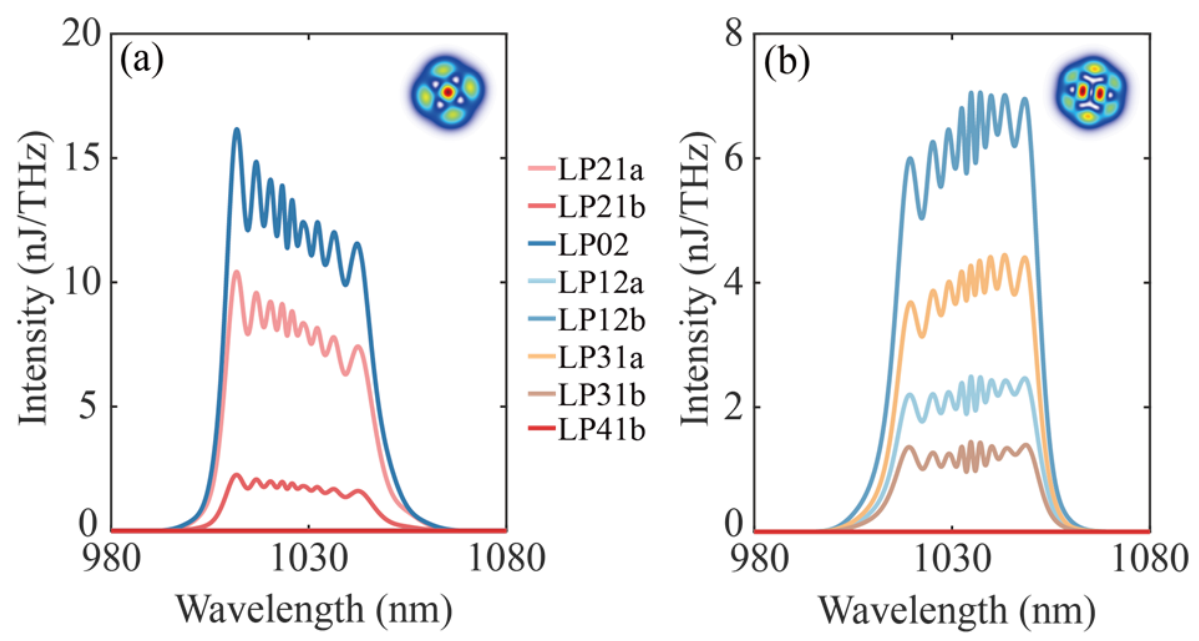

Fig. 9. Mode-resolved spectra of composite-mode solitons selectively generated in an MO based on 15-mode GIMFs: (a) an LP$_{02}$+LP$_{21}$ composite-mode soliton and (b) an LP$_{12}$+LP$_{31}$ composite-mode soliton. The insets show the corresponding total transverse intensity distributions. Modes with near-zero spectral intensities are not individually resolved in the figure, as their traces overlap with that of the LP$_{41b}$ mode.

## 4. CONCLUSION

In summary, we systematically investigate intermodal XPM dynamics inside multimode MOs and propose a spectrum-to-mode mapping mechanism for transverse-mode regulation of ultrafast pulses. Temporal overlap at the edges of pulses with different group velocities induces asymmetric spectral broadening and opposite spectral shifts through intergroup XPM. Offset intracavity spectral filters subsequently translate these spectral differences into differential loss, thereby preferentially retaining a group of modes with matched group velocities and achieving selective STML. This mechanism allows modal dispersion to serve as a resource for mode discrimination rather than as a detrimental effect requiring compensation.

In six-mode GIMF cavities, XPM-based seed preconditioning establishes an initial mode-group preference that guides mode selection by Filter1. Joint control of the filter detuning $\delta\lambda$ and saturation fluence $F_{\mathrm{sat}}$ then accesses distinct operating regimes supporting stable solitons with specific transverse-mode compositions, including the pure LP$_{01}$ mode, LP$_{11}$ degenerate modes, LP$_{21}$ degenerate modes, and LP$_{02}$+LP$_{21}$ composite modes. In regimes where Filter1 fails to discriminate modes, fast and slow pulses coexist in the MO and subsequently undergo intracavity collision driven by modal dispersion. By tailoring the temporal separation between the coexisting pulses, we show that steering the collision position relative to the two intracavity filters provides an additional route to govern mode selection. During soliton formation, intermodal FWM plays a dual role in regenerating modal components suppressed by the initial filtering and stabilizing specific relative phases among surviving modes. The latter role helps explain the consistent transverse-field structures observed for a given modal composition, notably yielding ultrafast $LG_{01}$ and $LG_{02}$ vortex pulses. Simulations in six-mode SIMFs and 15-mode GIMFs further support the applicability of the proposed framework across different refractive-index profiles, modal-dispersion characteristics, and expanded mode counts.

Finally, it is worth noting that the present scheme relies on external seed injection to initiate oscillation under large filter detuning, and solitons can only form from modes possessing closely matched group velocities. These constraints define the practical boundaries of our approach. Overall, spectrum-to-mode mapping provides a conceptually simple strategy for transverse mode manipulation in multimode fiber lasers, eliminating the requirement for additional spatial mode selective hardware. Controllable transverse-mode compositions lay a foundation for full spatiotemporal optical field engineering. Future efforts will target experimental demonstration of selective STML, extension toward higher-order composite-mode states, exploration of noise robustness and pulse energy scaling for structured light applications.

## APPENDIX A: SIMULATION EQUATION

The simulation of pulse propagation in the MMFs is based on the GMMNLSE with a saturable gain term:

$$\partial_z A_p(z,t) = \hat{D}A_p + i\frac{n_2\omega_0}{c}\left(1+\frac{i}{\omega_0}\partial_t\right)\hat{N}_p(A) + \hat{G}_p(A)$$

$$\hat{D} = i\delta\beta_p^{(0)} - \delta\beta_p^{(1)}\partial_t + \sum_{m=2}^{Nd} i^{m+1}\frac{\beta_p^{(m)}}{m!}\partial_t^m$$

$$\hat{N}_p(A) = \sum_{l,m,n}(1-f_R)S_{plmn}^K A_l A_m A_n^* + f_R S_{plmn}^R A_l \cdot \int_{-\infty}^{t} d\tau h_R(\tau) A_m(z,t-\tau) A_n^*(z,t-\tau)$$

$$\hat{G}_p(A) = \frac{g_0(\omega)}{2}A_p(t;z) - \sum_{l,m,n} S_{plmn}^G \partial_z A_n(t;z) \cdot \int \frac{A_l(t';z)A_m^*(t';z)}{F_{sat}}dt' \tag{A1}$$

where $\hat{D}, \hat{N}_p(A), \hat{G}_p(A)$ are linear, nonlinear, and gain operators, respectively. $\beta_p^{(j)}$, $j = 0, 1, \cdots, Nd$ is the $j$ th-order dispersion of mode $p$ obtained by performing a Taylor expansion of the propagation constant $\beta$. $\delta\beta_p^{(0)} = \beta_p^{(0)} - \beta_1^{(0)}$ represents the propagation-constant mismatch, whereas $\delta\beta_p^{(1)} = \beta_p^{(1)} - \beta_1^{(1)}$ represents the modal dispersion. $\frac{i}{\omega_0}\partial_t$ accounts for self-steepening. $f_R$ is the fractional Raman contribution, and $h_R(t)$ is the Raman response function. $S_{plmn}^K, S_{plmn}^R, S_{plmn}^G$ are mode overlap coefficients for the Kerr effect, Raman effect, and saturable gain effect, respectively. $g_0(\omega)$ is the small-signal gain spectrum, and $F_{sat}$ is the saturation fluence related to the doping of the gain medium in the gain fiber.

The simplified GMMNLSE reads:

$$\begin{aligned}\partial_z A_p(z,t) &= i\delta\beta_p^{(0)}A_p - \delta\beta_p^{(1)}\partial_t A_p \\ &+ i\frac{n_2\omega_0}{c}S_{pppp}^K|A_p^2|A_p \\ &+ i\frac{n_2\omega_0}{c}\sum_n^N S_{ppnn}^K A_p A_n A_n^*,\ p\neq n \\ &+ i\frac{n_2\omega_0}{c}\sum_n^N S_{pnpn}^K A_n A_p A_n^*,\ p\neq n\end{aligned} \tag{A2}$$

## APPENDIX B: MODAL ANALYSIS OF THE MULTIMODE FIBER

The simulation relies on MMFs supporting multiple modes, whose relevant parameters are obtained by solving the following waveguide field equation:

$$\nabla_t^2\psi(x,y) + k_0^2 n^2\psi(x,y) = \beta^2\psi(x,y) \tag{B1}$$

where $\nabla_t^2 = \frac{\partial^2}{\partial x^2} + \frac{\partial^2}{\partial y^2}$, $\psi(x,y)$ denotes the transverse modal field, $k_0 = \frac{2\pi}{\lambda_0}$ denotes the wave number of light waves in vacuum ($\lambda_0$ is the carrier wavelength), $n$ denotes the transverse refractive-index profile of the MMFs, and $\beta$ denotes the propagation constant of the field along the z-direction. Solving the eigenvalue problem in Eq. (B1) yields the guided modes, from which we calculate the linear parameters and nonlinear coupling coefficients shown in Fig. 10.

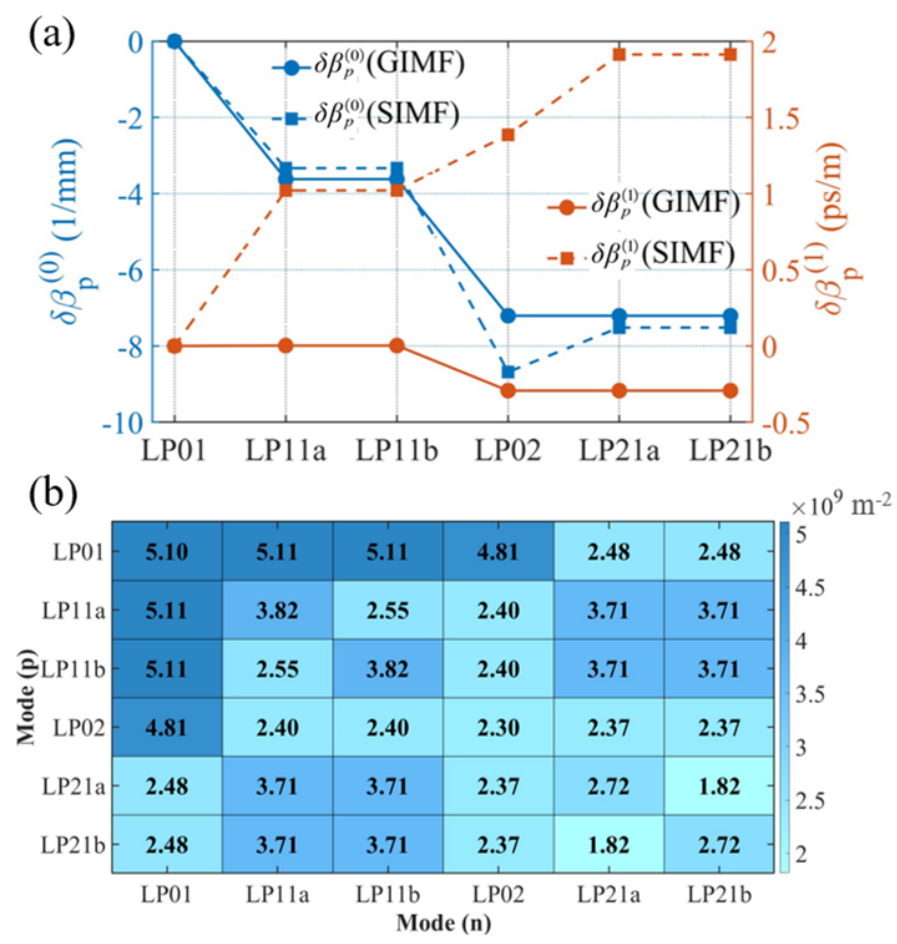

Fig. 10. (a) Propagation-constant mismatch and modal dispersion of the modes supported by GIMF and SIMF. (b) Nonlinear coupling coefficients for SPM (main diagonal elements, representing $S_{pppp}^K$ ) and XPM (off-diagonal elements, representing $S_{ppnn}^K + S_{pnpn}^K$) of the modes supported by GIMF.

In Fig. 10(a), the blue and orange markers represent $\delta\beta_p^{(0)}$ and $\delta\beta_p^{(1)}$, respectively. These data enable the evaluation of the FWM phase-matching conditions and the classification of the modes according to their group velocities. In Fig. 10(b), the diagonal elements correspond to the SPM nonlinear coupling coefficients $S_{pppp}^K$, whereas the off-diagonal elements represent the combined XPM coefficients $S_{ppnn}^K + S_{pnpn}^K$. As $S_{pppp}^K$ is inversely proportional to the effective mode area $A_{eff}$, a larger $S_{pppp}^K$ indicates a smaller $A_{eff}$.

## APPENDIX C: SEED GENERATION

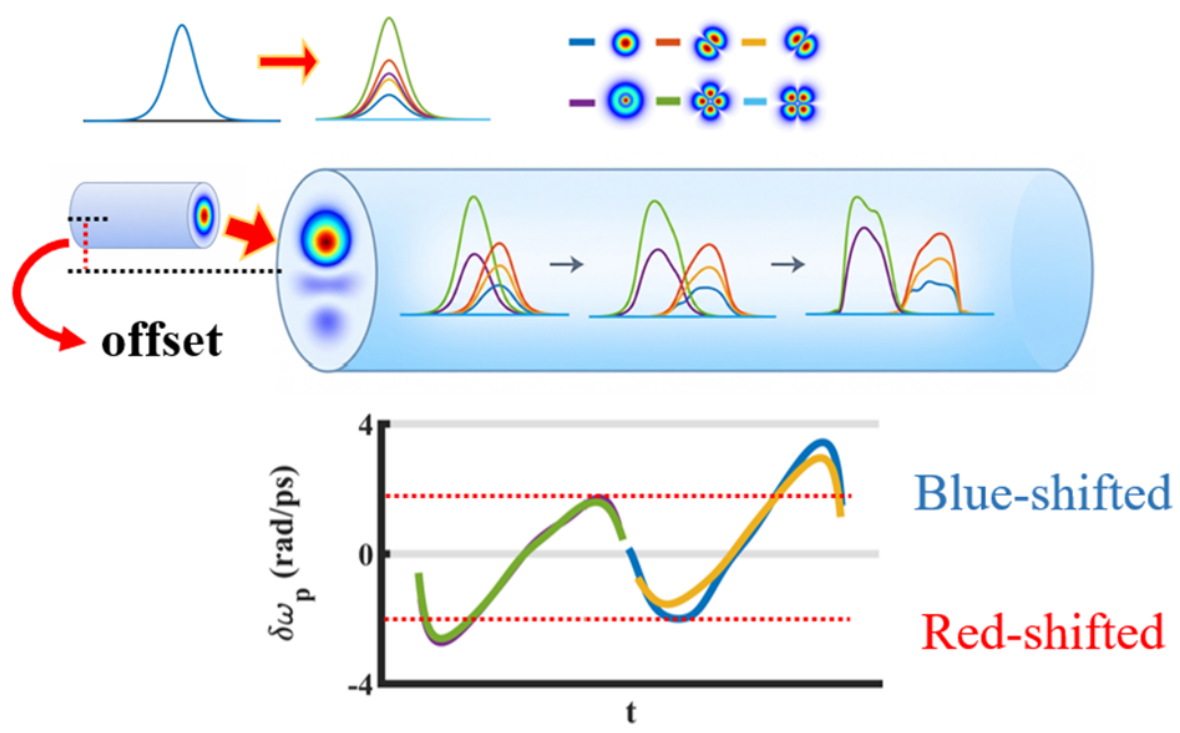

Fig. 11. Multimode-seed preprocessing by offset coupling and propagation through an MMF. The lower panel shows the instantaneous frequency shifts acquired by individual modes upon propagation.

In the simulations, we generate the multimode seed by offset-splicing an SMF to an MMF of the same type as the passive fiber within the MO. The mode-coupling transmission matrix is obtained by projecting the incident transverse field onto the eigenmode basis of the receiving fiber through mode matching [48]. The seed subsequently propagates through a preprocessing fiber segment of length $L_{\text{pre}}$, where modes with different group velocities undergo temporal walk-off due to modal dispersion. During this process, the modes pass through the temporal-overlap state illustrated in Fig. 2(e) and acquire instantaneous frequency shifts that can be exploited for mode selection, leading to the results shown in Fig. 4. Furthermore, varying the length of this fiber section controls the accumulated intermodal walk-off and hence the temporal separation between the modes, enabling the collision control demonstrated in Fig. 8.

**Funding.** National Natural Science Foundation of China (62405103).

**Disclosures.** The authors declare no conflicts of interest.

**Data Availability.** Data underlying the results presented in this paper are not publicly available at this time but may be obtained from the authors upon reasonable request.